# Earthquakes Three-Stage Early Warning and Short-Term Prediction

**Manana Kachakhidze[1,*], Nino Kachakhidze-Murphy[1], Giorgi Ramishvili [2], Badri Khvitia[3]**

1 Natural Hazard Scientific-Research Center, Georgian Technical University, Tbilisi 0175, Georgia
2 School of Natural Sciences and Medicine, Ilia State University, Sun and Solar System Department, Georgian Evgeni Kharadze National Astrophysical Observatory. Tbilisi 0162, Georgia
3 Sokhumi Institute of Physics and Technology, Tbilisi 0186, Georgia
* Correspondence: kachakhidzem@gmail.com

**Abstract:** The article discusses the possibilities of three-step early warning and short-term prediction of earthquakes based on the classical geological model of fault formation and a model of the generation of electromagnetic emissions detected before earthquakes. These possibilities are revealed by the analysis of frequency data of VLF/LF electromagnetic radiation, existing before the earthquake preparation period.

**Keywords:** earthquake; electromagnetic emissions; precursor.

## 1. Introduction

Professor Emeritus of the University of Tokyo and a member of Japan Academy, Seiya Uyeda wrote in 2013: „Japan's National Project for Earthquake Prediction has been conducted since 1965 without success. An earthquake prediction should be a short-term prediction based on observable physical phenomena or precursors. The main reason for no success is the failure to capture precursors.

Most of the financial resources and manpower of the National Project have been devoted to strengthening the seismographs networks, which are not generally effective for detecting precursors since many of precursors are non-seismic. The precursor research has never been supported appropriately because the project has always been run by a group of seismologists who, in the present author's view, are mainly interested in securing funds for seismology — on pretense of prediction. After the 1995 Kobe disaster, the project decided to give up short-term prediction and this decision has been further fortified by the 2011 M9 Tohoku Mega-quake. Thus, in Japan now, there is practically no support for short-term prediction research. Recently, however, substantial progress has been made in real short-term prediction by scientists of diverse disciplines [1].

We are not going to discuss the reasons that make it really impossible to describe the processes taking place in the focus of an earthquake from the point of view of seismology and lead to the fact that despite efforts aimed at strengthening the seismometer network, it is impossible to detect a seismic precursor to the earthquake.

To expand the study of the earhquake problem, since 2015 the Government of Japan has made significant investments in addressing the country's disproportionate earthquake and tsunami risk, informed by past lessons [2].

Despite this, only in the last year, earthquakes occurred in various seismically active countries of the world, which caused not only material damage but also huge human casualties (Turkey and Syria, on February 6, 2023, M7.8 and M7.5, Morocco, on September 8, 2023, M6.9, Philippines on November 17, 2023, M6.7, China, on December 18, 2023, M 6.2, Japan, on January 1, 2024, M7.6, etc.).

Modern ground-based and satellite observation methods have revealed multiple anomalous geophysical phenomena that accompany any natural disaster and are directly connected with the process of their preparation. Of course, this also concerns earthquakes. The complexity of the issue, despite the great scientific successes, gives us reason to think that, one country alone will not be able completely to solve the problems associated with any natural hazard.

Our group has been studying anomalous changes in various geophysical fields before an earthquake for a long time and concluded that for further research in the right direction, it is necessary to classify geophysical fields into triggers, indicators, and precursors according to their generation mechanisms and role in the earthquake preparation process [3].

This classification will simplify the vague picture created by the set of anomalous geophysical fields existent during the period of earthquake preparation that makes it difficult to detect the true precursor necessary for high-precision earthquake prediction.

Although scientific research on earthquake problems is widespread in the world, there are still many anomalous fields that exist before an earthquake, the research of which requires the implementation of various joint scientific projects involving researchers from different scientific areas of activity.

## 2. Discussion

### 2. 1. Some information on anomalous EM radiation before an earthquake.

According to Professor Emeritus Seiya Uyeda: "Earthquake prediction must specify the time, epicenter, and size of impending EQ with useful accuracy.

Among the long-, intermediate- and short-term predictions, only the short-term prediction is meaningful for directly protecting human lives and social infrastructures. The other two are mainly mere statistic forecasts based on past experiences and should not even be called prediction, although the intermediate-term forecast has entered into a new stage thanks to the GPS measurements" [1].

Our study found that VLF/LF EM emissions are responsible for earthquake short-term prediction [4-6].

We will return to this issue in detail below. Now we want to once again emphasize the properties of VLF/LF EM emissions, existing before earthquakes, which give it the possibility of short-term prediction.

A seismogenic zone since the time of transfer of perturbation along the system is not less than the oscillation period, can be considered a distributed system where the mass, elasticity (mechanical systems), capacity, and inductance (electric systems) elements are continuously spread in the whole volume of this system. This means that in the earthquake preparation area, each least element has its self-capacity and self-inductance because of piezoelectric, piezo-magnetic, electrochemical, and other effects.

Electromagnetic oscillations in the system are practically conditioned by constant and continuous changes in inductance and capacitance, which are caused by also constant changes in tectonic stress in the area of the earthquake focus.

Any change in capacitance (or inductance) in a virtual oscillatory circuit means that the circuit's natural frequency $\omega$ also changes instantly.

For its part, even the smallest change in the fault length is immediately reflected exactly on the change of capacitance (or inductance), i.e. in the change of the natural frequency ω of the circuit [4-6], which, during real observations in the earthquake preparation period, makes it possible to determine the change in the fault length (magnitude) in the focus with extremely high accuracy (1,2):

$$\omega = \beta \frac{c}{l} \qquad (1)$$

where ω is frequency of electromagnetic radiation, $l$ is the linear size of fault in earthquake focus, β is the characteristic coefficient of geological medium (it approximately equals 1).

$$\lg l = 0.6 Ms - 2.5 \qquad (2)$$

where *Ms* is earthquake magnitude [7].

It should be emphasized that the very high accuracy of measuring VLF/LF EM radiation also leads to a highly accurate determination of the rupture length. For example, prior to the L'Aquila catastrophic earthquake that occurred on 6 April 2009 besides Ultra-low frequency and kHz electromagnetic (EM) anomalies, 41MHz ($\omega_1$) and 54MHz ($\omega_2$) radiation were also recorded [8]. According to formula (1), the corresponding length of the emitting body of frequency $\omega_1$ would be $l_1 = 0.007$ km, and the corresponding length of $\omega_2$ would be $l_2 = 0.006$ km.

This means that it is possible to detect even the smallest changes in the length of microcracks by electromagnetic emission. However, there are restrictions in terms of reaching high-frequency electromagnetic radiation to the Earth's surface.

Our studies have proven, that for earthquake prediction, at the final stage of monitoring, it is crucial to determine the length of the rupture with an accuracy not of a kilometer, but of a higher order, which, as we mentioned above, is provided by EM radiation records.

Since, concerning earthquakes, electromagnetic radiation of the order of kHz is considerable (1≤M≤9 magnitude earthquakes correspond to frequencies of the spectrum in the range of 23 830 kHz≥f≥0.378 kHz), which is not absorbed by earth and if we consider the speed of electromagnetic waves, it can be considered that all the processes taking place in the focus, starting from the origin of microcracks to the final formation of the fault, can be observed in an almost parallel mode on the earth surface.

## 2. 2. Earthquake's early warnings and prediction

It is clear that expression "VLF/LF EM radiation is responsible for earthquake short-term prediction" means that in the case of VLF/LF EM emissions data monitoring, it should be possible quantitatively to describe the full cycle of earthquake preparation and occurrence based on the avalanche-like unstable geological model of fault formation [9] and a model of the generation of electromagnetic emissions detected before earthquakes [6].

A study was conducted for the Crete Earthquake (M = 5.6, 25/05/2016, 08:36:13 UTC) according to 73-day records of INFREP (the European Network of Electromagnetic Radiation).

Because the INFREP network fixes every minute amplitudes of 10 different frequencies of electromagnetic radiation it was calculated and used every minute frequency numerical value transformed by the normal distribution of Gauss [10].

At this stage, the cases of aftershocks and foreshocks are not considered.

According to the average hourly data of the 73-day frequency records of INFREP, by the formula (1) the lengths of the corresponding ruptures were calculated, and a graph (Figure1) was constructed, on which the initial moment of the avalanche-like unstable process of fault formation was clearly expressed.

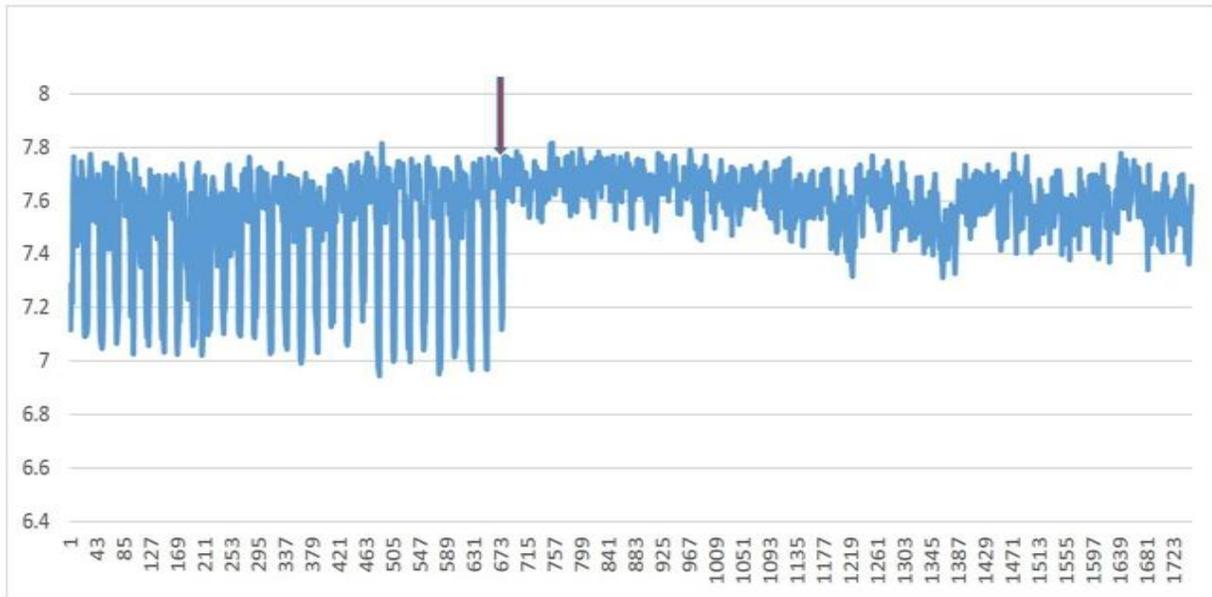

**Figure 1.** A graph of ruptures lengths by 73-day average hourly data. An arrow indicates the initial moment of the fault formation avalanche-like unstable process.

To analyze the fault length average hourly values data, we used the average square deviation method and calculated x ±σ significancies (Figure 2).

In real rocks, there always are randomly distributed defects (cracks of various sizes). Under the effect of tectonic stresses, it takes place a slow increase in the number and size of favorably oriented defects and the formation of new ones.

In a statistically homogeneous medium, under the effect of a uniformly distributed load, crack formation occurs throughout the entire volume. This quasi-homogeneous cracking corresponds to the subcritical stage of failure. One of the reasons for the cracking uniformity may be the formation of stable detachable cracks, as well as the stopping (delay) of cracks on the medium inhomogeneities. Already at this stage, the properties of the medium should change, for example, effective elastic moduli and quasi-anisotropy.

The total cycle of earthquake preparation and occurrence, which is depicted in Figure 2, is divided into stages:

**I Stage**. In the first stage, uniform cracking occurs in the earthquake preparation zone. During this period, only probabilistic conditions for the occurrence of an earthquake are created [9].

Research has found an important fact that even this early stage of earthquake preparation is accurately described by VLF/LF EM emissions data.

This was expected since even the slightest change in inductance and capacitance caused by crack formation is exactly reflected in the VLF/LF EM frequency data, which makes it possible to determine the length of the crack propagation zone formed under the influence of a uniformly distributed load. In this zone of crack propagation, only probabilistic conditions for the occurrence of an earthquake will be created in the future (Figure 2).

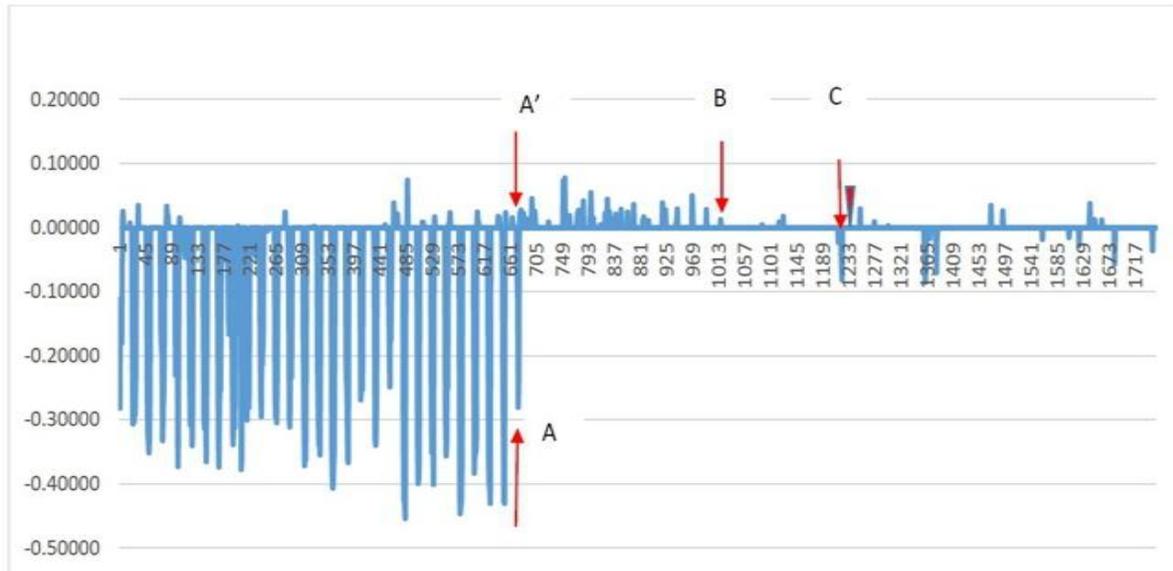

**Figure 2.** A graph portrayal of the total cycle of earthquake preparation and occurrence, constructed by the x ±σ significances of the hourly values of the rupture lengths. The avalanche period of fault origin is marked between A, A' and B. The so-called period of "silence" is marked between B and C, the triangle marks the moment of the earthquake occurrence.

For example, the average length of fracture formations from the beginning of data records to point A equals 7.118 *km* (here and everywhere below INFREP minute records are used for the values of rupture lengths and the time moments).

It has also been revealed that long before an earthquake, at the very beginning of the fault formation avalanche process, it is possible to "mark" the approximate length of the rupture of an impending earthquake (for example, at point A' in Figure 2), that is the preliminary determination of the magnitude.

In our considered case, the "marked-probable" length of the rupture of the impending earthquake was approximately 7.73 *km*, and it was already possible to determine it 50 days before the earthquake.

However, it should be noted here that during the total cycle of earthquake preparation, even according to the geological model, the length of this probable main fault of the impending earthquake adjusts.

**II Stage.** The transition to the second stage occurs when the average density of ruptures reaches a certain critical value in the entire volume or a significant part of it. As a result of the interaction of cracks, an avalanche stage of a given earthquake preparation sets in.

The involvement of an increasing number of inhibited cracks in this process is associated with a rapid and sharp redistribution of the local stress field due to the cracks combining a higher rank (larger size).

An avalanche increase in the number and sizes of cracks leads to a sharp increase in the rate of general deformation and a change in the integral physical characteristics of the medium.

If this process leads to an earthquake, then it must be unstable [9].

Since one of the important stages of earthquake preparation starts from the avalanche process, to describe this stage, let us use the excerpt from Figure 2 (Figure 3). Our studies also confirmed that this stage is indeed an active stage of fault formation (Figure 3. AA'B). Consider this process in more detail:

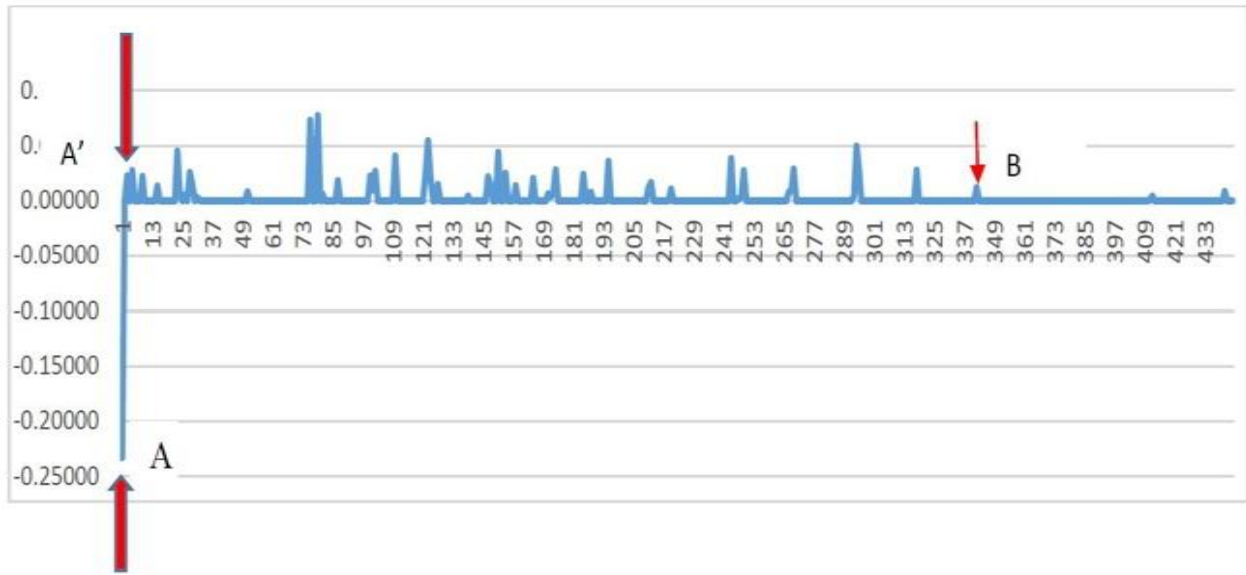

**Figure 3.** The active stage of the avalanche-unstable process of fault formation.

It was found that the beginning of the avalanche-unstable process is distinctly manifested in a sharp change in the length of the cracks already formed by this time. In the case we considered, this sharp change in the length of the cracks started at 3:00 on May 2 and continued until 6:00 on the same day. During these approximately three hours, as expected from the geological model, it appears that a rapid coalescence of cracks did indeed occur. The length of united cracks changed from 7.118 *km* to 7.73 *km*. (section A A' in Figure 3).

During this active stage of the avalanche-unstable process, the main fault of the future earthquake is intensively growing, and by the end of this stage, the length of the fault is almost completely formed. This fact is pictorially reflected in Figure 3, expressed by positive anomalies in the AA' B section. It is obvious, that the process is not continuous, there are small "silence" periods between anomalies (intervals of time during which the fault length hardly changes). In consideration case, the average duration of these small "silences", calculated in hours, is 8.4, and the maximum duration is 24. This process continues until the beginning of a much longer main "silence" (Figure 3, point B), after which the process moves to the third stage of the formation of the main fault. It happened at 15 May 9:00. At this moment the fault length was 7.828 *km*. These studies allow us to declare as the early warning first stage of an impending earthquake the moment of such a sharp change onset in the length of cracks in a relatively short time, which in our case begins at 6 o'clock on May 2 (Figure 3. Point A'). In our particular case, this active stage of the formation of fault lasted for 13 days and 3 hours. During this period, the length of the fault increased to 7.828 *km*. It should be noted here that this first stage of notification only indicates that an earthquake of this particular magnitude is indeed expected at this particular location, however, before the earthquake, we have some time until the main fault of the future earthquake is finally formed (at least 10-14 days). It is clear, that the duration of this time will depend on the geological particularities of the environment. After the first notification, there are still stages to go through described by the classical, qualitative geological model, which, during the monitoring period, should be accurately reflected in VLF/LF EM radiation records.

**III Stage.** In the third stage, the increase in deformation is already accompanied by a drop in stress. Due to the heterogeneity of the properties of the medium, the unstable deformation is drawn up into a narrow zone in which several relatively large cracks form. At the same time, due to the general drop in average macro stresses in most of the volume, the cracks stop developing and partially close up. As a result, the integral deformation rate of the total zone decreases at this third stage (instability stage). Take place the restoration of many integral characteristics of the zone. A narrow zone of unstable deformation is characterized by an increased concentration of cracks and represents itself as the surface of the future main fault [9]. This process, in fact, is the so-called main "silence" period, which begins immediately after the end of the active stage of the avalanche process and graphically is represented by the BC area in Figure 4. It is clear, that minor changes in the main fault length of the impending earthquake are expected even during the main "silence" period.

The third, i.e. the main stage of "silence" can be considered starting from point B (Figures 3, 4), the duration of which should already be approximately 3 times greater than the average "silence" value of the previous stage (no doubt, this value will depend on the characteristics of the region).

In the case of the earthquake we discussed, the main "silence" period began on 15 May at 09:00 and ended on 24 May at 16:00, i.e. it lasted for 9 days and 7 hours.

At the moment of the onset of the main "silence", the second stage of early warning about the upcoming earthquake can already be announced, which will only adjust the time of the earthquake occurrence. In particular, it is indicated that before the earthquake less and less time left. It means that we are waiting for an alarm about an incoming earthquake.

The main "silence" phase ends, where the very first pre-earthquake anomaly appears (Figure 4, point C'). In our case, during the main "silence" the length of the fault decreases from 7.828 *km* to 7.5 *km*.

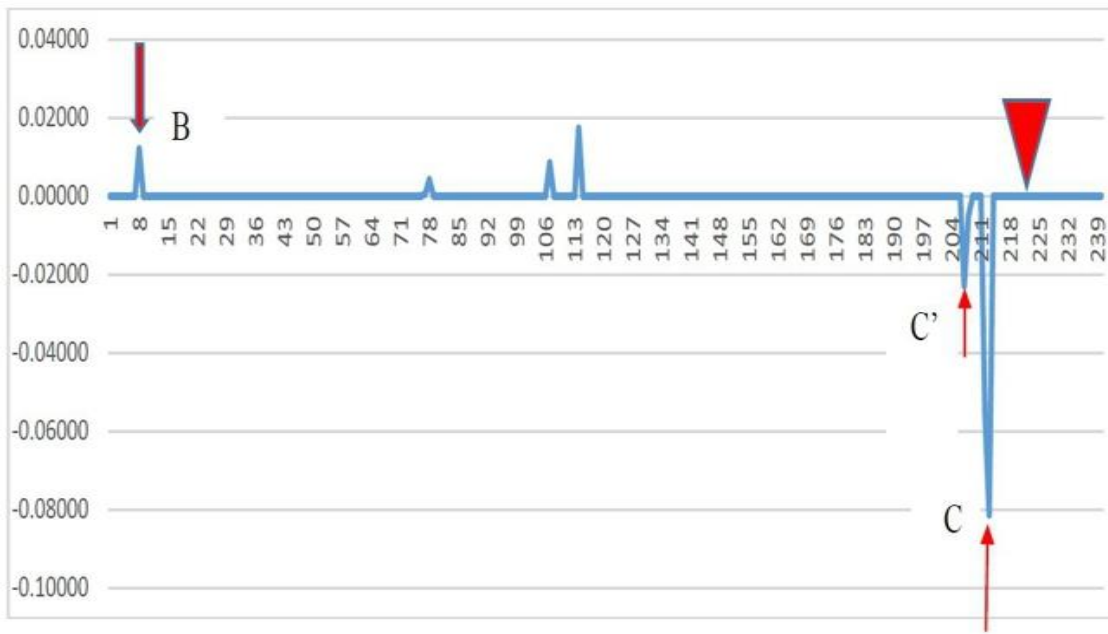

Figure 4. (Excerpt from Figure 2). The main "silence" period with the two anomalies preceding the earthquake (C' and C). An earthquake is marked with a red triangle.

As known, the fault - an earthquake is formed by ripping apart bridges between large cracks.

The process of ripping apart bridges qualitatively is similar to the total process, and therefore it must be preceded by a short-term and smaller amplitude change in the deformation rate.

Since the destruction of one of the bridges may not be sufficient to rip apart the entire main fault, there may be several such short-term changes in the deformation rate. In this case, at least some of these fluctuations must precede large foreshocks, and the last of them - the immediate foreshock or the main shock itself [9].

The real earthquake records accurately describe the qualitative geological model process described above. In particular, immediately after the end of the main "silence", two negative anomalies were revealed, which indicate the closing of microcracks and the reduction of the rupture length (Figures 4,5).

Let's consider the period after the end of the main "silence" in detail (Figure 5):

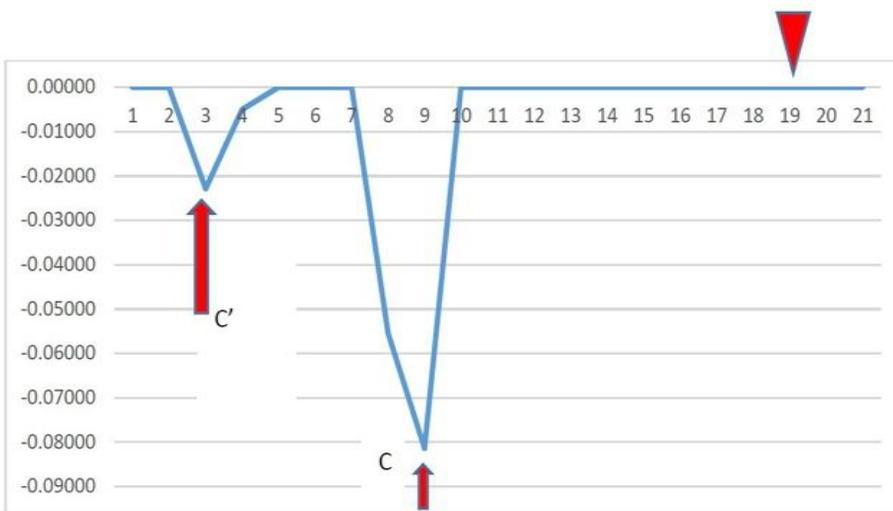

Figure 5. (Excerpt from Figure 2). Post-silence anomalies (C' and C) and the moment of the earthquake occurrence (red triangle).

The analysis showed that the first destruction of one of the bridges occurred on May 24 at about 16:00 and at that time the length of the rupture was 7.5 *km* (the corresponding anomaly is marked with C' in Figures 4 and 5).

As it was expected from the geological qualitative model, this destruction turned out to be insufficient to rip up the total main fault.

The second destruction of another bridge occurred on May 24 at 22:00, when the length of the rupture became 7.496 *km*.

The final short-term change in the rate of deformation that caused the earthquake occurred on May 25 at 08:36, 16 hours and 36 minutes after the appearance of the first anomaly and 10 hours and 36 minutes after the appearance of the second anomaly. At the moment of the earthquake, the length of the main fault has been reached 7.475 *km*.

Thus, based on the above said, it can be concluded that at the moment of detecting the anomalies (anomaly) that occur immediately after the end of the main "silence", it is already possible to announce an alarm about an incoming earthquake, because at this time, besides the magnitude and location of the impending earthquake [6], we already know that the earthquake is only a few hours away.

### 2. 3.  The additional practical use of the research

It is noteworthy that with the continuous extension of coal mining deeper underground, ground stress and gas pressure in coal seams gradually increase, and coal and gas outbursts, rock bursts, and other coal rock dynamic disasters become more serious and complex [11].

It is known that in the processing of rocks in the mines electromagnetic radiation occurs. In this direction, relevant studies have been going on for a long time [12-15].

Landslides also are one of the catastrophic events that result in massive destruction and loss of lives. Hence, an appropriate technique is essential to predict potential weak slip planes which may eventually lead to landslides. Recently, a study appeared, where Fracture Induced Electromagnetic Radiation (FEMR) technique was used to identify such districts of potential "activity" [15, 16].

Research, similar to the above proposed by us can make a significant contribution to the study of the stress-deformed state of rock mass around tunnels and underground structures in terms of perfecting assessment methods, as well as identifying landslide-prone areas by detecting the adjoining weak slip planes.

### 3. Conclusions

Based on the conducted studies, it was found that the description of the qualitative avalanche-unstable geological model of fault formation by VLF/LF EM radiation data made it possible to quantitatively characterize the complete cycle of earthquake preparation and occurrence:

(1) It is possible approximately to determine the length (magnitude) of the main rupture of the incoming earthquake several tens of days before the earthquake.

(2) In the case of monitoring, the beginning of the "avalanche process" of ruptures is observed, during which there is a sharp, rapid increase in the length of the cracks. The beginning of the "avalanche process" can already be considered as the first stage of early warning of an incoming earthquake.

(3) Immediately after the end of the active avalanche process, the so-called "silence" period begins during which a second stage of early warning of an incoming earthquake can be announced.

(4) "Silence" can be considered finished at the moment when the first anomaly, indicating a radical change in the length of the main fault, appears. At this time, it already is possible to announce an alarm about an incoming earthquake.

(5) The time interval between the first and second stages of early warning of an incoming earthquake allows us to minimize human and material damage caused by an earthquake.

(6) The possibilities due to the quantitative description of earthquake preparation and occurrence process have been created on the base classical geological qualitative model of fault formation.

(7) In the case of monitoring, based on the frequency data of VLF/LF electromagnetic radiation before the earthquake, the methods of earthquake three-stage early warning (including alarm at the last stage) and earthquake short-term prediction have been created.

(8) The offered research ensures the safest conditions for the development of deep underground mineral deposits and the prediction of landslides with high accuracy.

**Author Contributions**

**Conflicts of Interest**

**Acknowledgments**

The authors are grateful to the network INFREP for providing us with electromagnetic emissions data used in this paper.